\documentclass[12pt]{iopart}
\usepackage{amsfonts}
\usepackage{iopams}
\usepackage{amssymb}
\usepackage{graphicx}
\usepackage{bm}
\usepackage{cite}
\usepackage[colorlinks=true,linkcolor=blue,anchorcolor=blue,citecolor=blue,urlcolor=blue]{hyperref}
\expandafter\let\csname equation*\endcsname\relax
\expandafter\let\csname endequation*\endcsname\relax
\usepackage{amsmath}

\begin{document}

\title{Elementary excitations in a spin-orbit-coupled spin-1 Bose-Einstein condensate}

\author{Yuanyuan Chen$^{1}$, Hao Lyu$^{2}$, Yong Xu$^{3}$, and Yongping Zhang$^{1}$}
\address{$^{1}$Department of Physics, Shanghai University, Shanghai 200444, China}
\address{$^{2}$Quantum Systems Unit, Okinawa Institute of Science and Technology Graduate University, Onna, Okinawa 904-0495, Japan}
\address{$^{3}$Center for Quantum Information, IIIS, Tsinghua University, Beijing 100084, China}

\ead{yongxuphy@tsinghua.edu.cn and yongping11@t.shu.edu.cn}

\begin{abstract}
While a spin-orbit-coupled spin-1 Bose-Einstein condensate has been experimentally observed,
its elementary excitations remain unclear in the stripe phase.
Here, we systematically study the elementary excitations in three distinct phases of a spin-orbit-coupled spin-1 Bose-Einstein condensate.
We find that the excitation spectrum as well as the corresponding static response function and structure factor
depend strongly on spin-orbit coupling parameters such as the quadratic Zeeman field and the Rabi frequency. In the stripe phase, besides two gapless Goldstone modes, we show the existence of roton excitations. Finally, we demonstrate that quantum phase transitions
between these different phases including the zero-momentum, plane wave and stripe phases
are characterized by the sound velocities and the quantum depletion.
\end{abstract}


\submitto{\NJP}

\maketitle

\section{Introduction}

Experimental realizations of spin-orbit-coupling in ultracold neutral atomic gases~\cite{Lin,Ji2014,Wang,Cheuk,Bromley,Burdick,WangZY} open an avenue for studying
exotic quantum phases and nonlinear dynamics~\cite{Wu,Khamehchi2017,LiCH,Kroeze,WuR,Zhang2019}
or simulating topological phases~\cite{MengZM,ZhangDW,Cooper2019,YongXu2019}.
In neutral atoms, the so-called spins are pseudo-spins which may come from hyperfine states. The minimum system to support spin-orbit coupling is a spin-1/2 one which involves two hyperfine states. These hyperfine states are coupled by
two oppositely propagating Raman lasers via a two-photon transition so that the momentum exchange
between the Raman lasers and two hyperfine states generates atomic Raman-induced spin-orbit coupling.
In such a system,
intriguing ground-state phases for Bose-Einstein condensates (BECs) including the stripe, plane-wave and zero-momentum phases have been revealed. Quantum phase transitions between these three phases depend on spin-orbit coupling parameters and interaction strengths~\cite{Martone,LiYun2012,LiYun2013}. The stripe phase, which spontaneously breaks both the gauge symmetry and continuous translational symmetry, has been identified as possessing supersolid properties~\cite{Lyu,LiJR,Bersano,LiS,Sanchez-Baena}. The plane-wave phase spontaneously breaks the time-reversal symmetry and features nonzero magnetizations,
and the zero-momentum phase is analogous to a conventional BEC without spin-orbit coupling.
Such exotic BECs support intriguing elementary excitations: roton excitations in the plane-wave phase and two
gapless Goldstone excitations in the stripe phase~\cite{Martone,LiYun2013,ChenXL}.
The roton excitations have been experimentally observed in the plane-wave phase~\cite{Ji,Khamehchi}.

In stark contrast to a solid-state system,
spin-orbit coupling can be synthesized in a spin-1 cold atom system consisting of three hyperfine states.
In fact, spin-orbit coupled spin-1 BECs have been realized in a recent experiment using three
Raman lasers~\cite{Campbell}. The BEC exhibits a richer ground-state phase diagram due to spinor interactions and tunable quadratic Zeeman shifts~\cite{Campbell,Lan,Natu,Yu,Sun,Martone2016,Cabedo,Saha}. It has also been shown that there are multiple
rotons in elementary excitations of the spin-1 plane-wave phase~\cite{Yu,Sun}.
However, elementary excitations in the stripe phase of a spin-1 BEC have not yet been investigated (possibly due to the fact that
it is much harder to study these excitations), and their fundamental properties such as Goldstone modes are still unknown.

In this work, we systematically study elementary excitations in a spin-orbit-coupled spin-1 BEC. Exotic ground states and phase transitions in this system make elementary excitations more interesting. In the stripe ground state, the existence of density modulations gives rise to a Bloch band-gap structure for elementary excitations. Such an excitation Bloch spectrum has two gapless Goldstone modes
associated with the spontaneous breaking of two different continuous symmetries.
We also find roton excitations at finite quasimomenta in the stripe phase, which is a unique feature of the
stripe phase of a spin-1 BEC; such roton excitations have not been found in the stripe phase of a spin-1/2 BEC.
In addition, we use the static response function and structure factor to characterize the elementary excitations.
Both the static response function and structure factor diverge at a particular quasimomentum in the stripe phase,
which is a salient feature of a stripe ground state. In the plane-wave ground state, apart from gapless excitations,
there appear multiple rotons. The static response function presents sharp peaks around the location of rotons when
a roton excitation is sufficiently softened. From the linear dispersion of the lowest excitation in all these phases,
we can determine the sound velocities. We find that the sound velocities and the quantum depletion can be used
as signals to reflect quantum phase transitions between different ground states.

This paper is organized as follows. In Sec.~\ref{model}, we describe our theoretical frame for analyzing elementary excitations, response functions and quantum depletion. In Sec.~\ref{results}, we show  numerical results of elementary excitation for both spin-1 antiferromagnetic and  ferromagnetic BECs. The conclusion is finally presented in Sec.~\ref{conclusion}.

\section{Theoretical model}
\label{model}

We consider a three-dimensional spin-1 spinor BEC with a Raman-induced spin-orbit coupling.  Such spin-orbit coupling can be realized by applying three Raman lasers along the $x$ direction to atoms with hyperfine states $m_F=0,\pm1$. These lasers induce a two-photon Raman transition between two hyperfine states $|m_F=0\rangle$ and $|m_F=+1\rangle$ ($|m_F=-1\rangle$). The Hamiltonian of the system is
\begin{align}
\hat{H}=\int d\bm{r}\hat{\Psi}^\dagger \hat{H}_\mathrm{SOC}\hat{\Psi}
+\frac{1}{2}\int d \bm{r} \Psi^\dagger
\left[g_0 \hat{\Psi}^\dagger\hat{\Psi} + g_2 \hat{\Psi}^\dagger\hat{\bm{F}}\hat{\Psi} \cdot \hat{\bm{F}} \right]\hat{\Psi}.
\label{Hamiltonian}
\end{align}
Here, $\hat{\Psi}=(\hat{\Psi}_1,\hat{\Psi}_2,\hat{\Psi}_3)^T$ represent the bosonic annihilation operators of the atoms with different pseudo-spins, and $\hat{\bm{F}}=(\hat{F}_x,\hat{F}_y,\hat{F}_z)$ are the spin-1 Pauli matrices. $\hat{H}_\mathrm{SOC}$ is the single-particle spin-orbit-coupled Hamiltonian,
\begin{equation}
\hat{H}_\mathrm{SOC}=\frac{1}{2} \left(-i\frac{\partial}{\partial x}-2 \hat{F}_z\right)^2+\frac{1}{2}\nabla^2_\perp+\Omega \hat{F}_x + \epsilon \hat{F}^2_z,
\label{singleparticle}
\end{equation}
where $\nabla_\perp=\partial^2_y+\partial^2_z$, $\Omega$ is the Rabi frequency of the Raman lasers which depends on the laser intensity,
and $\epsilon$ is the quadratic Zeeman shift which can be tuned by controlling the detuning of the lasers~\cite{Campbell}. In Eq.~(\ref{Hamiltonian}) and the following calculations, we set all quantities dimensionless. The units of the momentum, length, density, and energy are $\hbar k_\mathrm{Ram}$, $1/k_\mathrm{Ram}$, $k^3_\mathrm{Ram}$, and $\hbar^2k^2_\mathrm{Ram}/m$, respectively, where $m$ is the atom mass and $k_\mathrm{Ram}=2\pi/\lambda_\mathrm{Ram}$ is the wave vector of the Raman lasers with $\lambda_\mathrm{Ram}$ being the wavelength. $g_0$ and $g_2$ are spin-independent and spin-dependent interaction strengths, respectively, which depend on the atom number and the $s$-wave scattering lengths in the total spin channels~\cite{Stamper-Kurn}. In this work, the spin-independent interaction is always assumed to be repulsive ($g_0>0$), and the spin-dependent interaction could be either antiferromagnetic ($g_2>0$) or ferromagnetic ($g_2<0$).

In the mean-field approximation, the perturbations can be omitted and we have $\langle \hat{\Psi}_\sigma \rangle=\psi_\sigma$. The mean-field energy functional can be written as
\begin{equation}
\mathcal{E}=\int d\bm{r}\psi^\ast \hat{H}_\mathrm{SOC}\psi +\frac{1}{2}\int d \bm{r} \psi^\ast
\left[g_0 |\psi|^2+ g_2 \psi^\ast\hat{\bm{F}}\psi \cdot \hat{\bm{F}} \right]\psi.
\label{energy}
\end{equation}
By minimizing the energy functional, we obtain the ground-state wave functions.
Once we know the ground state, we can study elementary excitations of the system. The bosonic field operators can be written as the sum of the ground-state wave functions and fluctuation operators,
\begin{equation}
\hat{\Psi}_\sigma=e^{-i\mu t}\psi_\sigma + e^{-i\mu t}\delta\hat{\Psi}_\sigma,
\label{operator}
\end{equation}
with
\begin{equation}
\delta\hat{\Psi}_\sigma=\sum_{\bm{q},l} \left[u^{(l)}_{\bm{q},\sigma}(\bm{r})e^{-i\omega_l t}\hat{a}_{\bm{q},l} +v^{(l)\ast}_{\bm{q},\sigma}(\bm{r})e^{i\omega_l t}\hat{a}^\dagger_{\bm{q},l} \right].
\end{equation}
Here, ${\bm{q}}$ is the quasimomentum of the excitation, $\omega_l$ is the excitation energy with $l$ being the label of the $l$th band of the excitation spectrum, and $\mu$ is the chemical potential. $u^{(l)}_{\bm{q},\sigma}(\bm{r})$ and $v^{(l)}_{\bm{q},\sigma}(\bm{r})$ ($\sigma=1,2,3$) are the perturbation amplitudes, which satisfy the normalization condition,
\begin{equation}
\sum_{\sigma} \int d\bm{r} \left[|u^{(l)}_{\bm{q},\sigma}(\bm{r})|^2-|v^{(l)}_{\bm{q},\sigma}(\bm{r})|^2 \right]=1.
\label{renormalization}
\end{equation}
$\hat{a}_{\bm{q},l}$ and $\hat{a}^\dagger_{\bm{q},l}$ are the annihilation and creation operators of quasiparticles of the $l$th excitation band, respectively. Then we write the Heisenberg motion equation for the operator $\hat{\Psi}_\sigma$,
\begin{equation}
i\frac{\partial\hat{\Psi}_\sigma}{\partial t}=[\hat{\Psi}_\sigma,\hat{H}].
\end{equation}
By substituting Eqs.~(\ref{Hamiltonian}) and (\ref{operator}) into the above equation, we obtain the equations for $\psi$ and $\delta\hat{\Psi}$.
The equations for the ground-state wave functions are
\begin{equation}
\mu\psi=\left(\hat{H}_\mathrm{SOC} +\hat{H}_\mathrm{int}[\psi] \right)\psi,
\end{equation}
with
$$\hat{H}_\mathrm{int}[\psi]=g_0 |\psi|^2+ g_2 \psi^\ast\hat{\bm{F}}\psi \cdot \hat{\bm{F}}.$$
The perturbation amplitudes $u^{(l)}_{\bm{q},\sigma}$ and $v^{(l)}_{\bm{q},\sigma}$ satisfy the Bogoliubov-de Gennes (BdG) equations~\cite{Dalfovo},
\begin{equation}
\mathcal{L}\phi=\omega_l \phi,
\label{BdG}
\end{equation}
with $\phi=(u^{(l)}_{\bm{q},1},u^{(l)}_{\bm{q},2},u^{(l)}_{\bm{q},3},v^{(l)}_{\bm{q},1},v^{(l)}_{\bm{q},2},v^{(l)}_{\bm{q},3})^T$. The matrix $\mathcal{L}$ is
\begin{equation}
\mathcal{L}=
\left(
\begin{array}{cc}
H_\mathrm{SOC}(\bm{q})+X-\mu & Y \\
-Y^\ast & -H_\mathrm{SOC}(-\bm{q})-X^\ast+\mu
\end{array}
\right),
\end{equation}
with
\begin{align}
X&=X_1+(g_0+g_2)X_2+2g_2X_3,\nonumber\\
Y&=(g_0+g_2)Y_1+g_2Y_2+Y_3.
\end{align}
The matrices $X_j$ and $Y_j$ ($j=1,2,3$) are
\begin{align}
X_1&=\left(
\begin{matrix}
h1 & 0 & (g_0-g_2)\psi_1\psi^\ast_3 \\
0 & h_2 & 0 \\
(g_0-g_2)\psi^\ast_1\psi_3 & 0 & h_3
\end{matrix}
\right),
X_2=\left(
\begin{matrix}
0 & \psi_1\psi^\ast_2 & 0 \\
\psi^\ast_1\psi_2 & 0 & \psi_2\psi^\ast_3  \\
0 & \psi^\ast_2\psi_3 & 0
\end{matrix}
\right), \nonumber\\
X_3&=\left(
\begin{matrix}
0 & \psi_2\psi^\ast_3 & 0 \\
\psi^\ast_2\psi_3 & 0 & \psi_1\psi^\ast_2 \\
0 & \psi^\ast_1\psi_2 & 0
\end{matrix}
\right),
\end{align}
and
\begin{align}
Y_1&=\left(
\begin{matrix}
\psi^2_1 & \psi_1\psi_2  & 0 \\
\psi_1\psi_2  & 0 & \psi_2\psi_3 \\
0 & \psi_2\psi_3 & \psi^2_3
\end{matrix}
\right),
Y_2=\left(
\begin{matrix}
0 & 0  & \psi^2_2 \\
0  & 2\psi_1\psi_3 & 0 \\
\psi^2_2 & 0 & 0
\end{matrix}
\right), \notag\\
Y_3&=\left(
\begin{matrix}
0 & 0  & (g_0-g_2)\psi_1\psi_3 \\
0  & g_0 \psi^2_2& 0 \\
(g_0-g_2)\psi_1\psi_3 & 0 & 0
\end{matrix}
\right),
\end{align}
with
\begin{eqnarray}
h_1&=&2(g_0+g_2)|\psi_1|^2+(g_0+g_2)|\psi_2|^2+(g_0-g_2)|\psi_3|^2,\nonumber\\
h_2&=&2g_0|\psi_2|^2+(g_0+g_2)(|\psi_1|^2+|\psi_3|^2), \nonumber\\
h_3&=&(g_0-g_2)|\psi_1|^2+(g_0+g_2)|\psi_2|^2+2(g_0+g_2)|\psi_3|^2. \nonumber
\end{eqnarray}

Atom density fluctuations can be induced when perturbations are applied on the system, which can be characterized by the density response function. In the linear response theory, the static density response function can be written as~\cite{LiYun2013,Pitaevskii,Blakie}
\begin{equation}
\chi(\bm{q})=\frac{1}{N} \sum_l\frac{1}{\omega_l}
\left|\sum_\sigma \int d\bm{r}e^{-i\bm{q} \cdot \bm{r} } \left[\psi_{\sigma}^* (\bm{r} )   u^{(l)}_{\bm{q},\sigma}(\bm{r}) +\psi_{\sigma} (\bm{r} ) v^{(l)}_{\bm{q},\sigma}(\bm{r}) \right]  \right|^2,
\end{equation}
where $N$ is the total atom number. 
Here, the summation should include both the positive and negative excitation bands obtained by solving Eq.~(\ref{BdG}).

Experimentally, excitations in a BEC can be probed by Bragg spectroscopy~\cite{Ji,Stenger,Petter}. The Bragg beams transfer momentum and energy to the system. The excitation probability of Bragg scattering with momentum $\bm{q}$ can be described by the static structure factor~\cite{LiYun2013},
\begin{equation}
S(\bm{q}) =
\frac{1}{N}\sum_{\omega_l>0}
\left| \sum_\sigma \int d\bm{r}e^{-i\bm{q} \cdot \bm{r} }  \left[\psi_{\sigma}^* (\bm{r} )   u^{(l)}_{\bm{q},\sigma}(\bm{r}) +\psi_{\sigma} (\bm{r} ) v^{(l)}_{\bm{q},\sigma}(\bm{r}) \right]   \right|^2,
\end{equation}
which can be treated as the sum of the excitation probabilities to different excitation bands. Static structure factor of a spin-orbit-coupled spin-1/2 BEC has been studied theoretically in Refs.~\cite{Martone,LiYun2013}, and experimental measurement in the plane-wave phase has also been reported~\cite{Ji}.

The above theoretical frame belongs to the mean-field theory, which is valid when quantum fluctuations are negligible. The quantum fluctuations induce a fraction of the condensation depleted even at zero temperature. The quantum depletion in a spin-orbit-coupled BEC can be calculated if the excitation spectrum and amplitudes are known~\cite{ChenXL,Zheng,Cui}. The density of depleted atoms can be calculated as~\cite{ChenXL}
\begin{equation}
n_\mathrm{ex}=\frac{1}{V}\sum_{\sigma,l}\int d\bm{q}\int d\bm{r} |v^{(l)}_{\bm{q},\sigma}( \bm{r} )|^2,
\end{equation}
where $V$ is the volume of the system, and the perturbation amplitude $v^{(l)}_{\bm{q},\sigma}$ depends on the quasimomentum $\bm{q}$.  If the number of depleted atoms is much smaller than the total atom number, then the mean-field approximation is valid.

There are plane-wave, zero-momentum and stripe phases for the gound state of a spin-orbit-coupled spin-1 BEC~\cite{Yu,Sun}, which have different excitation properties. In the following, we will describe the theoretical treatments for different phases.

\subsection{Plane-wave and zero-momentum phases}

In the plane-wave or zero-momentum phase, the system has a uniform density profile. The ground-state wave functions of these two phases
can be assumed as
\begin{equation}
\psi=\left(
\begin{array}{c}
\psi_1 \\ \psi_2 \\ \psi_3
\end{array}
\right)=\sqrt{n}e^{ikx}\left(
\begin{array}{c}
\varphi_1e^{i\theta_1} \\ \varphi_2 \\ \varphi_3e^{i\theta_3}
\end{array}
\right),
\end{equation}
where $n=N/V$ is the atom density, and $k$ is the quasimomentum. If $k\ne 0$, then the system is in the plane-wave phase, and if $k= 0$, then the system is in the zero-momentum phase. The wave functions satisfy the normalization condition, $\varphi^2_1+\varphi^2_2+\varphi^2_3=1$. With the above wave functions,  the energy functional in Eq.~(\ref{energy})  becomes
\begin{align}
\frac{\mathcal{E}}{N}&=\frac{k^2}{2} +2k  (\varphi_1^2 -\varphi_3^2) + \left(\epsilon+2 \right)  (\varphi^2_1 + \varphi^2_3) \nonumber\\
&\phantom{={}}+ 2\Omega\varphi_2(\varphi_1\cos\theta_1+\varphi_3\cos\theta_3)+\frac{g_0n}{2}
+\frac{g_2n}{2}(1-2\varphi^2_3)^2 \nonumber\\
&\phantom{={}}+\frac{g_2n\varphi^2_2}{2}\left[4\varphi_1\varphi_3\cos(\theta_1+\theta_3)+4\varphi^2_3-\varphi^2_2\right].
\end{align}
The parameters $k$, $\varphi_{1,2,3}$ and $\theta_{1,3}$ can be determined by minimizing $\mathcal{E}$. The excitation spectrum from Eq.~(\ref{BdG}) can be calculated by using the ansatz of plane-wave excitations~\cite{Yu}
\begin{align}
u^{(l)}_{\bm{q},\sigma}(\bm{r})&=U^{(l)}_{\bm{q},\sigma} e^{ikx+i\bm{q}\cdot \bm{r}},\nonumber\\
v^{(l)}_{\bm{q},\sigma}(\bm{r})&=V^{(l)}_{\bm{q},\sigma} e^{ikx+i\bm{q}\cdot \bm{r}},
\end{align}
where  $U^{(l)}_{\bm{q},\sigma}$ and  $V^{(l)}_{\bm{q},\sigma}$ are plane-wave amplitudes.
The excitation spectrum of the plane-wave phase exhibits phonon modes in the long-wavelength limit and roton modes at a finite momentum, which will be discussed in Sec.~\ref{results}. We will also show that the excitation spectrum can be tuned by changing spin-orbit coupling parameters.

\subsection{Stripe phase}

In the plane-wave phase, roton instability emerges when the roton energy becomes zeros or even negative by tuning relevant parameters. In this situation, the plane wave is no longer the ground state of the system. The ground state turns out to be the stripe phase with a spatially modulated density profile~\cite{LiYun2013}.

Wave functions of the stripe phase can be constructed as the superposition of plane waves with different wave numbers,
\begin{equation}
\psi=\left(
\begin{array}{c}
\psi_1 \\ \psi_2 \\ \psi_3
\end{array}
\right)=\sqrt{n}\sum^{L}_{j=-L} e^{ijKx}
\left(
\begin{array}{c}
\varphi^{(j)}_1 \\ \varphi^{(j)}_2 \\ \varphi^{(j)}_3
\end{array}
\right),
\label{stripe}
\end{equation}
with the normalization condition
$$\sum_{\sigma,j}|\varphi^{(j)}_\sigma|^2=1.$$
$L$ is the cutoff of the plane-wave modes, and $K$ determines the period of the stripes. $\varphi^{(j)}_\sigma$ and $K$ are determined by substituting Eq.~(\ref{stripe}) into Eq.~(\ref{energy}) followed by minimizing the resultant energy functional. This procedure shows that only coefficients of plane waves with $j$ being an odd number (i.e., $j=\pm1,\pm3,\cdots$.) are nonzero. We find that the period of the density is $\pi/K$, which is similar to that of the spin-1/2 case~\cite{LiYun2013,ChenXL}.

With the ground states, we can calculate the elementary excitation spectrum. Since the wave functions of the stripe phase is periodic, the BdG equations is also periodic and has the same period. Therefore, the perturbation amplitudes $u^{(l)}_{\bm{q},\sigma}(\bm{r})$ and $v^{(l)}_{\bm{q},\sigma}(\bm{r})$ are Bloch waves, which can be assumed to be
\begin{align}
u^{(l)}_{\bm{q},\sigma}(\bm{r})&=e^{i\bm{q}\cdot \bm{r}}\sum^{L}_{j=-L}U^{(l,j)}_{\bm{q},\sigma}e^{i(2j+1)Kx}, \nonumber\\
v^{(l)}_{\bm{q},\sigma}(\bm{r})&=e^{i\bm{q}\cdot \bm{r}}\sum^{L}_{j=-L}V^{(l,j)}_{\bm{q},\sigma}e^{i(2j+1)Kx}.
\label{stripe-pertubation}
\end{align}
Here, $\bm{q}$ is the quasimomentum and the superscript $l$ indicates the $l$th band. By substituting Eq.~(\ref{stripe-pertubation}) into the BdG equations Eq.~(\ref{BdG}), we obtain the excitation spectra and the corresponding excited states, from which the response function and structure factor can be computed.

\section{Results and discussions}
\label{results}

Now we are at the stage to study elementary excitations in the spin-orbit-coupled spin-1 BEC. In the following, we will study both the antiferromagnetic BEC ($g_2>0$) and the ferromagnetic BEC ($g_2<0$).

\begin{figure*}[htbp]
\centering
\includegraphics[width=16cm]{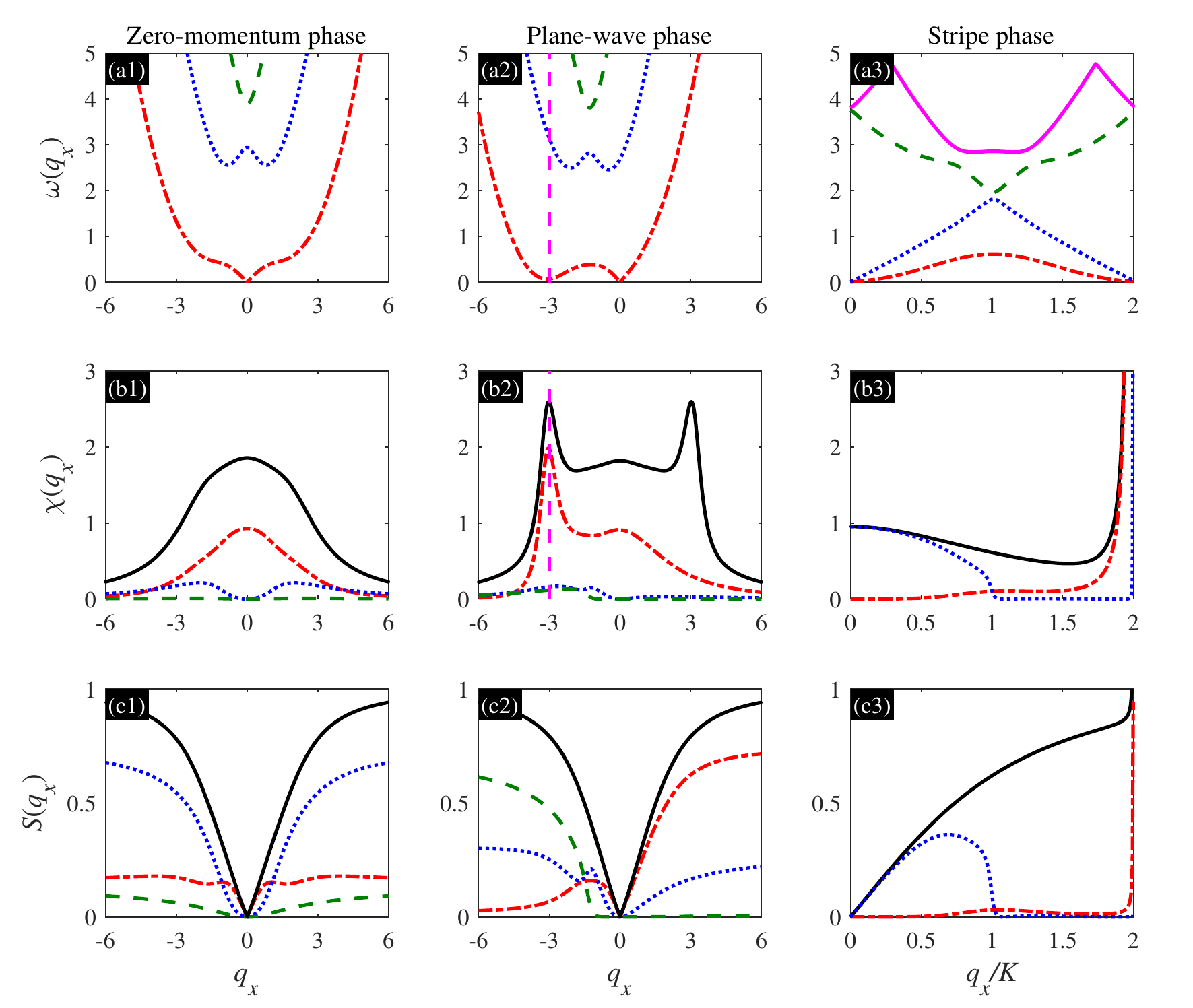}
\caption{(a1)-(a3) Excitation spectrum, (b1)-(b3) static response function and (c1)-(c3) static structure factor of a 
spin-orbit-coupled spin-1 BEC with an antiferromagnetic interaction. 
The first, second and third columns correspond to the zero-momentum phase, plane-wave phase and stripe phase, respectively.
In (a1)-(a3), different lines represent different bands. 
In (a1) and (a2), there are three bands, and in (a3), only four lowest bands are plotted.
In (a2), the vertical magenta-dashed line marks out the location of the roton at $q_x=-3$. 
In (b1) and (b2), the red-dot-dashed, blue-dotted, and green-dashed lines represent response functions of the three bands, 
and the black-solid line denotes the total response function obtained by adding together
the contributions of all positive and negative bands.  
In (b3), the red-dot-dashed and blue-dotted lines refer to the static response functions contributed by the lowest two gappless bands, and the black-solid line refers to the total static response function obtained by adding together
the contributions of all bands in (a3). In (c1) and (c2), the red-dot-dashed, blue-dotted, and green-dashed lines 
describe the structure factor of different bands; their sum gives rise to the total structure factor denoted by the black-solid line. In (c3), the red-dot-dashed and blue-dotted lines describe the structure factor of the lowest two bands, and the black-solid line 
describes the total one. $\pi/K$ [$K$ is defined in Eq.~(\ref{stripe})] denotes the period of the stripe phase. The quadratic Zeeman shift is set as $\epsilon=-0.2,-0.7$, and $-1.5$ in the first, second, and third columns, respectively.
Here, $\Omega=1.8$, $g_0n=1$ and $g_2n=0.1$.
}
\label{Fig1}
\end{figure*}

\subsection{Antiferromagnetic BEC}

The spin-orbit-coupled antiferromagnetic gas can undergo phase transitions between the the stripe phase, the plane-wave phase and the zero-momentum phase, which depend on the quadratic Zeeman shift and the Rabi frequency. 
It is also found that the stripe phase occupying the $\left|m_F=\pm1\right\rangle$ states can exist in a large parameter region~\cite{Yu,Campbell}, 
which makes it experimentally accessible to study its excitation properties. Based on previous works on the ground states of such a system~\cite{Yu,Sun,Martone2017}, we present our results on elementary excitations in the following.

In Fig.~\ref{Fig1}, we show the excitation spectra of the three distinct phases (corresponding to different quadratic Zeeman shifts $\epsilon$) and their corresponding static response functions and structure factors.
Here, the Rabi frequency and interaction strength are fixed at $\Omega=1.8$, $g_0n=1$ and $g_2n=0.1$, respectively. 
The three different phases can emerge by tuning $\epsilon$~\cite{Yu}. Since the excitations along the $y$ and $z$ directions are irrelevant to spin-orbit coupling and thus have similar results as the conventional BEC, we here only consider the excitations along the $x$ direction with spin-orbit coupling and set $q_y=q_z=0$. When $\epsilon=-0.2$, the ground state is the zero-momentum phase with $k=0$. Our calculation shows the existence of three branches of excitation spectra, which is symmetric respect to the quasimomentum $q_x$, as shown in Fig.~\ref{Fig1}(a1). The static response function $\chi(q_x)$ has a single local maximum at $q_x=0$, which corresponds to the ground state [see the black line Fig.~\ref{Fig1}(b1)]. Here, the red-dot-dashed, blue-dotted, and green-dashed lines represent the contributions from the first, second, and third excitation bands in Fig.~\ref{Fig1}(a1), respectively. 
We note that the negative-band contributions to $\chi(q_x)$ are not shown here since we find that the response function of $l$th band possess the symmetry
$\chi_l(q_x)=\chi_{-l}(-q_x)$, where $\chi_{\pm l}(q_x)$ correspond to the energy $\pm\omega_{l}(q_x)$.
Similarly, the static structure factor $S(q_x)$ is plotted in Fig.~\ref{Fig1}(c1) with contributions from different excitation bands.
The black-solid line refers to the total structure factor obtained by adding together the contributions of the three positive bands. We see that the three lines are symmetric with respect to $q_x$ arising from the symmetric excitation spectra.

\begin{figure}[ht]
	\centering
	\includegraphics[width=15cm]{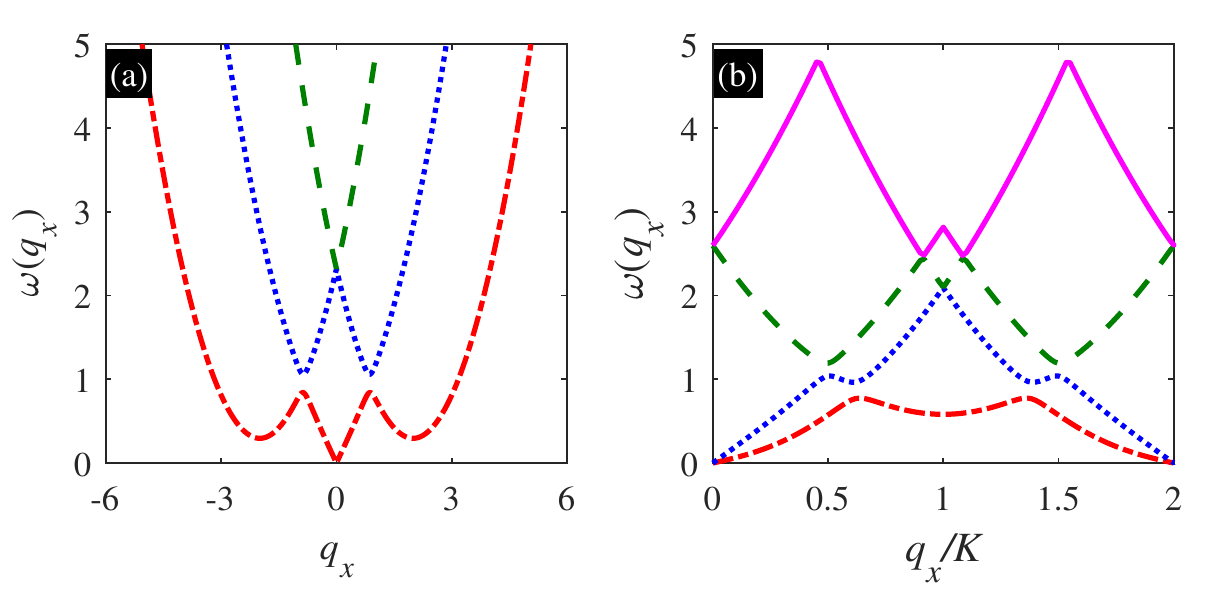}
	\caption{Excitation spectrum of the zero-momentum phase (a) and the stripe phase (b) in the antiferromagnetic case. 
		In (a), there are three bands, and in (b), only four lowest bands are shown [different colored lines represent different bands].
		In (a), the quadratic Zeeman shift $\epsilon=0.2$, and in (b), $\epsilon=-0.2$. 
		Here, $\Omega=0.2$, $g_0n=1$ and $g_2n=0.1$.}
	\label{Fig2}
\end{figure}

When we change $\epsilon$ to $-0.7$, the system is in the plane wave phase.
Figure~\ref{Fig1}(a2) shows the existence of one linear excitation
and a roton excitation, which is consistent with previous results~\cite{Martone,Yu,Sun}.
The roton excitation is a part of the lowest branch of the excitation spectrum, featuring a local minimum around a finite quasimomentum $q_\mathrm{rot}\approx -3$, the location of which is marked out by the vertical magenta-dashed line. The minimum of the roton energy is so low that two peaks around $q_x=\pm q_\mathrm{rot}$ emerge in the response function [see Fig.~\ref{Fig1}(b2)]. We note that the peak on the right side originates from unphysical negative-band contributions, which are not shown here. Different from the structure factor in the zero-momentum phase, the contributions of the three excitation bands are asymmetric; however, the total $S(q_x)$ (the black-solid line) is always symmetric with respect to $q_x=0$. We also see that the contributions of the first and second bands have a local maximum, which corresponds to the local maximum of the corresponding excitation band. We also note that the structure of the static structure factor for the plane-wave phase in the spin-1 case is vary similar to that of the spin-1/2 case, which has been measured in a recent experiment~\cite{Ji}.

In Fig.~\ref{Fig1}(a2), the low roton energy leads to two peaks in the curve of $\chi(q_x)$, which are higher than the peak at $q_x=0$. If we further decrease $\epsilon$, we expect the occurrence of the roton instability induced by roton softening and the divergence of the response function.
When this happens, the ground state becomes the stripe phase with spatially modulated density profile instead of the plane-wave phase.
Because of the density pattern, the BdG equation in Eq.~(\ref{BdG}) is periodic with the same period $\pi/K$, leading to the excitation spectrum featuring Bloch band-gap structures as shown in Fig.~\ref{Fig1}(a3).
In stark contrast to the zero-momentum and plane wave phases with one gapless mode, in the stripe phase, there are two gapless Goldstone modes arising from simultaneously spontaneous breaking of the gauge symmetry and the continuous translational symmetry~\cite{LiYun2013}.
The response function and the structure factor are shown in Figs.~\ref{Fig1}(b3) and \ref{Fig1}(c3), respectively. At $q_x=2K$, we find $\sum_{\sigma,j}|U^{(l,j)}_{\bm{q},\sigma}|^2-|V^{(l,j)}_{\bm{q},\sigma}|^2=0$ for $l=1,2$, so that the normalization condition for $u^{(l)}_{\bm{q},\sigma}$ and $v^{(l)}_{\bm{q},\sigma}$ in Eq.~(\ref{renormalization}) is invalid. This leads to the divergence of $\chi(q_x)$ and $S(q_x)$ at $q_x=2K$. A similar divergence has been reported in a spin-1/2 BEC~\cite{LiYun2013}. 
The divergence of the static structure factor at the boundary of the Brillouin zone is a character of the spatially modulated density profile.

More interesting phenomena can be observed when we choose a small Rabi frequency so that the dispersion of the single-particle spin-orbit-coupled Hamiltonian $\hat{H}_\mathrm{SOC}$ has a three-well shape in the lowest branch. Figure~\ref{Fig2} shows the excitation spectrum of the zero-momentum phase and the stripe phase for $\Omega=0.2$ and different $\epsilon$. When $\epsilon=0.2$, the ground state of the system is the zero-momentum phase. In the excitation spectrum, there emerge two symmetric rotons, which is very different from Fig.~\ref{Fig1}(a1) where rotons do not exist. We note that by increasing $\Omega$, the two rotons stiffen with enlarged roton gaps, leading to the disappearance of the rotons for a sufficiently large $\Omega$. In this case, the ground state occupies the middle well of the three symmetric wells in the single-particle dispersion~\cite{Yu}. Rotons are excited at the positions corresponding to the other two wells.
If we tune $\epsilon$ to a negative value, the ground state becomes the stripe phase~\cite{Yu}. There, compared with Fig.~\ref{Fig1}(a3), two gapless modes still exist in the excitation spectrum; interestingly, we also find roton excitations in the two lowest bands [see Fig.~\ref{Fig2}(b)]. Due to the periodicity of Bloch bands, these rotons are distributed periodically in the repeated Brillouin zone representation, reminiscent of the roton excitation of the spin-orbit coupled BEC in optical lattices~\cite{Martone-OL, Toniolo}. In the latter case, rotons originates from the ground state occupying only one Bloch state with a nonzero quasimomentum~\cite{Martone-OL, Toniolo}. Here, the emergence of roton structures in periodic excitation bands can be understood with the assistance of the single-particle dispersion from the Eq.~(\ref{singleparticle}). In this parameter regime, the single-particle dispersion have two symmetric wells located at $k_{\pm}$ and a well at $k=0$ with a higher energy. The ground state occupies the two symmetric wells, leading to the stripe phase~\cite{Yu}. If  atoms can be excited to the middle well, roton structures emerge in the excitation spectrum.

\begin{figure}[ht]	
\centering
\includegraphics[width=15cm]{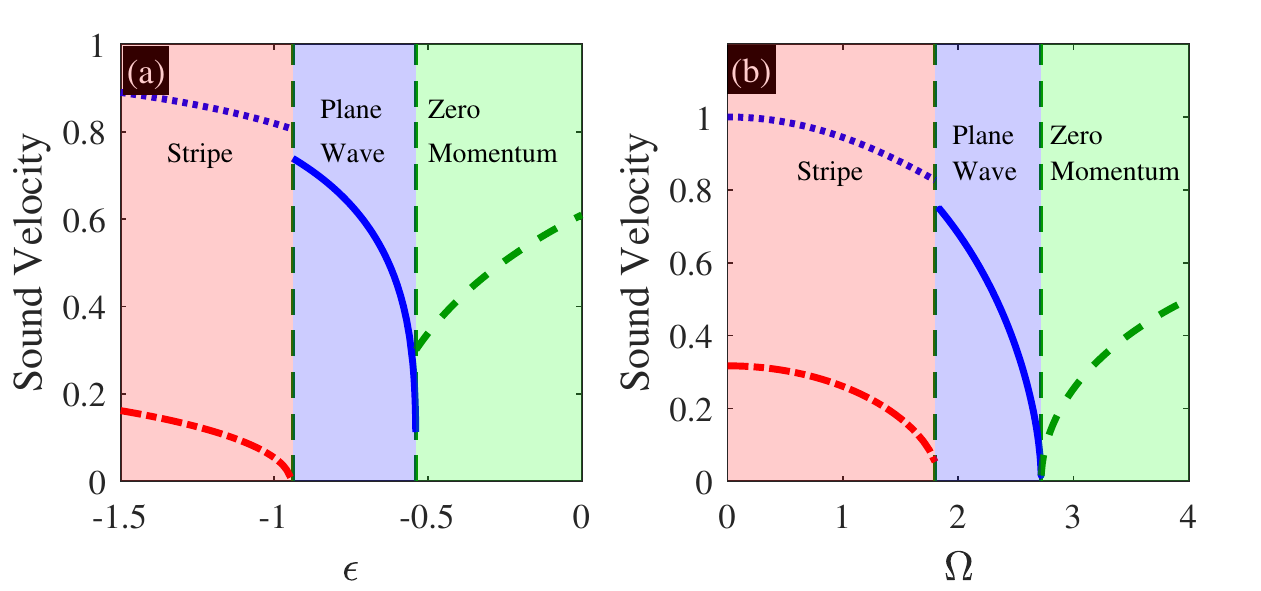}
\caption{The sound velocity as a function of (a) the quadratic Zeeman shift $\epsilon$ ($\Omega=1.8$) and (b) 
the Rabi frequency $\Omega$ ($\epsilon=-1$). The different colored backgrounds represent different phases of the ground state. 
In the stripe phase, there are two distinct sound velocities since the excitation spectrum has two different gapless modes.  }
\label{Fig3}
\end{figure}

Based on the fact that under the long wavelength limit, the lowest elementary excitations are always gapless with linear dispersion
and the excitation spectrum depends strongly on spin-orbit coupling parameters,
we expect that the phase transitions between these different phases can be characterized by the sound velocities.
Figure~\ref{Fig3}(a) plots the sound velocities with respect to the quadratic Zeeman shift $\epsilon$. When $\epsilon>-0.5$,
the ground state is the zero-momentum phase, and the sound velocity declines as we decrease $\epsilon$.
It exhibits discontinuity across the phase transition point around $\epsilon\approx-0.5$:
In the plane wave phase, the sound velocity approaches zero whereas in the zero-momentum phase, it approaches a finite value.
The discontinuity of the sound velocity indicates that the transition between the zero-momentum and plane-wave phases driven by $\epsilon$ is a first order
phase transition, consistent with the previous results~\cite{Yu}.
The sound velocity in the plane-wave phase rises as $\epsilon$ further decreases. At $\epsilon\approx-0.8$, the system enters into the stripe phase with two gapless Goldstone modes in the excitation spectrum, resulting in two different sound velocities [see Fig.~\ref{Fig3}(a)].
The discontinuity of the sound velocity at the transition point indicates that the transition between the plane wave phase
and the strip phase is of the first order. In addition, we display the sound velocity with respect to the Rabi frequency in Fig.~\ref{Fig3}(b). When we increase $\Omega$ from a very small value, we see that the system changes from the stripe phase
to the plane-wave phase and finally to the zero-momentum phase. Similar to the case shown in Fig.~\ref{Fig3}(a),
the sound velocity experiences a discontinuity across the stripe and plane wave phases, implying that the phase transition
is of the first order. Yet, different from the preceding case, the sound velocities approach zero as $\Omega$ approaches the transition point from either the plane wave phase or the zero-momentum phase, indicating that the transition between the plane wave and
zero-momentum phases
is of the second order, which is in agreement with previous results in Ref.~\cite{Yu}.

\begin{figure}[htbp]	
	\centering
	\includegraphics[width=9cm]{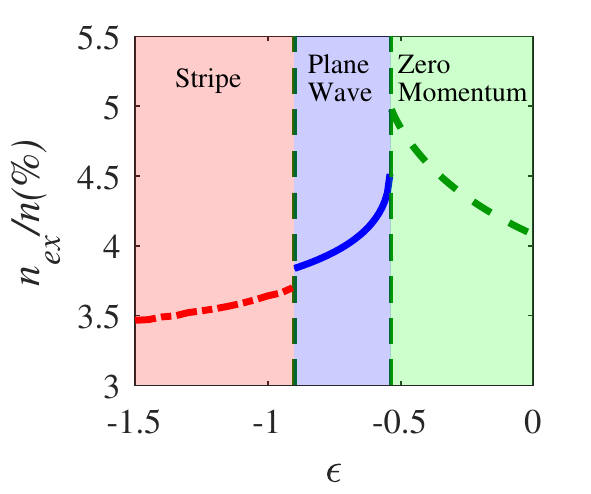}
	\caption{The quantum depletion $n_\mathrm{ex}/n$ with respect to the quadratic Zeeman shift $\epsilon$ ($\Omega=1.8$).
		The different colored backgrounds represent different phases of the ground state. Here, $g_0n=1$ and $g_2n=0.1$ 
		with $n=1.0k^3_\mathrm{Ram}$, which is a standard experimental density.   }
	\label{Fig4}
\end{figure}

We also plot the quantum depletion $n_\mathrm{ex}$ as a function of $\epsilon$ in Fig.~\ref{Fig4} [where the same parameters are chosen as in Fig.~\ref{Fig3}(a)]. The quantum depletion in the stripe and plane-wave phases rises with $\epsilon$ increasing, while it declines with $\epsilon$ increasing in the zero-momentum phase, which are similar to those in the spin-1/2 case~\cite{ChenXL}.
The first order phase transitions between these three distinct phases have also been revealed by the discontinuous
change of the quantum depletion. We note that in our parameter regimes, $n_{\mathrm{ex}}/n$ is less than 5\%, which verifies the validity of the mean-field treatments of our system.

\begin{figure*}[htbp]
	\centering
	\includegraphics[width=16cm]{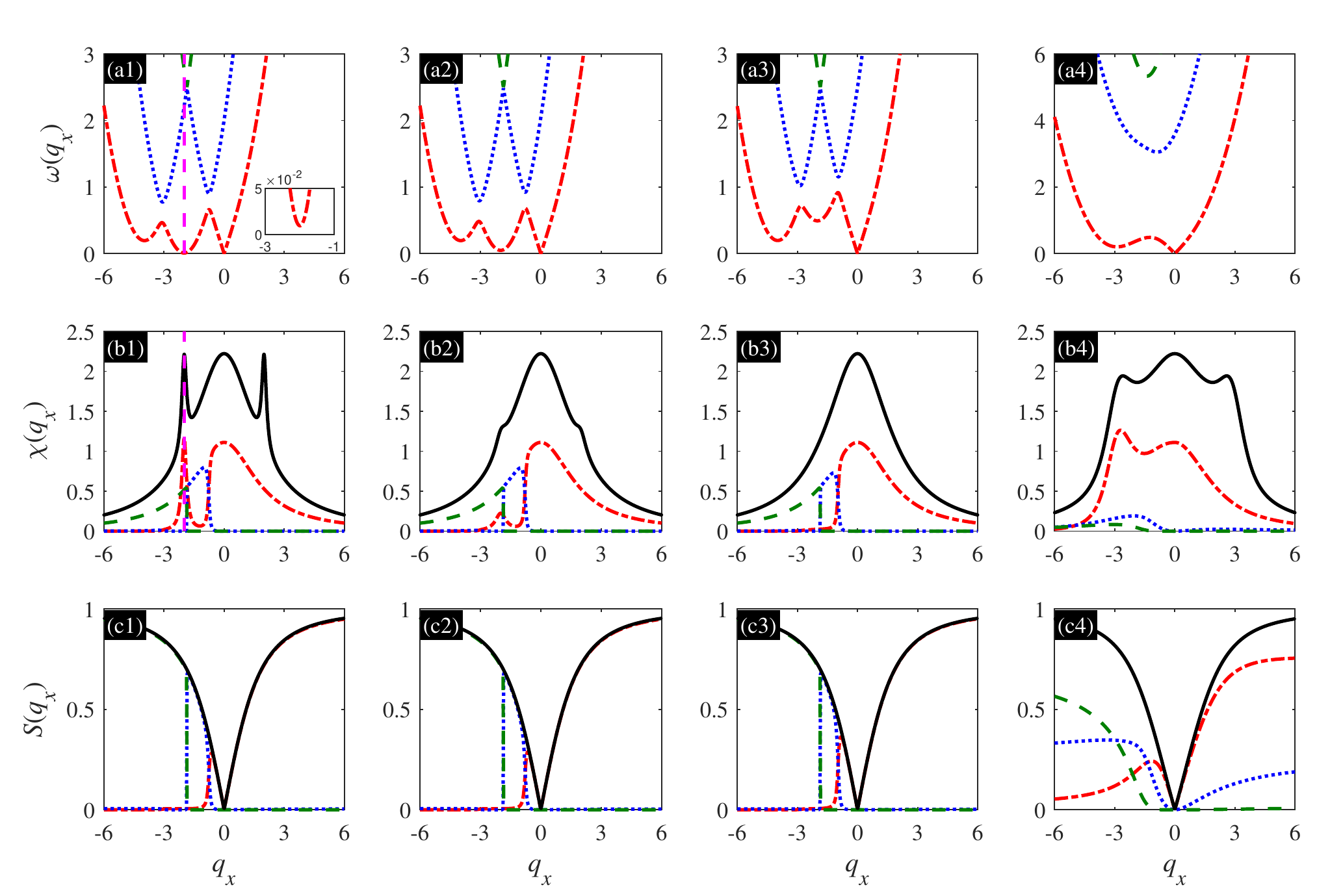}
	\caption{(a1)-(a4) Excitation spectrum, (b1)-(b4) static response function and (c1)-(c4) static structure factor of a 
		spin-orbit-coupled spin-1 BEC with a ferromagnetic interaction. 
		In (a1), the vertical magenta-dashed line marks out the location of the roton, and the zoomed-in view of the roton is shown in the inset. 
		Similar to Fig.~\ref{Fig1}, we use different colored lines to represent different bands in (a1)-(a4).
		In (b1)-(b4) and (c1)-(c4), the red-dot-dashed, blue-dotted and green-dashed lines depict the response function and the structure factor associated with the different bands in (a1)-(a4); the black-solid lines are the total response function and the total structure factor. 
		In the first column, $\epsilon=-0.01$ and $\Omega=0.2$; in the second column, $\epsilon=-0.05$ and $\Omega=0.2$; in the third column, $\epsilon=-0.5$ and $\Omega=0.2$; in the forth column, $\epsilon=-1.5$ and $\Omega=2.5$. We also set $g_0n=1$ and $g_2n=-0.1$.}
	\label{Fig5}
\end{figure*}

\subsection{Ferromagnetic interaction}

Now, we turn to study the ferromagnetic case with $g_2<0$. 
In this scenario, the stripe phase can only exist within a small parameter region, which is very hard to be observed in experiments~\cite{Campbell,Yu}. 
We also find that it is very difficult to obtain the accurate ground-state wave function for the stripe phase in the parameter regime, 
rendering the calculated excitation spectrum unreliable. 
We thus only consider the elementary excitations of the plane-wave and zero-momentum ground states.

In Fig.~\ref{Fig5}, we plot the elementary excitation spectra of the ferromagnetic gas and its corresponding response functions and structure factors. For $\epsilon=-0.01$, a double-roton structure emerges in the excitation spectrum [see Fig.~\ref{Fig5}(a1)], as a result of the triple-well shape in the dispersion of $\hat{H}_\mathrm{SOC}$. The roton at $q_x\approx-4$ has a higher energy and the roton at $q_x\approx-2$ has a lower energy. 
The energy of the latter roton is so low that two peaks at $q_x\approx\pm 2$ emerges in the response function, as shown in Fig.~\ref{Fig5}(b1). In addition, the two maxons in the excitation spectrum induce a maximum 
in the curves of the first-band contribution to the structure factor [see Fig.~\ref{Fig5}(c1)]. By decreasing $\epsilon$, we find that the roton at $q_x\approx-4$ softens and the roton at $q_x\approx-2$ stiffens [see Fig.~\ref{Fig5}(a2) and (a3)]. Consequently, the two peaks at $q_x\approx\pm 2$ in $\chi(q_x)$ disappear. If we choose a large negative $\epsilon$ and a large $\Omega$, the dispersion of the single-particle spin-orbit-coupled Hamiltonian $\hat{H}_\mathrm{SOC}$ has a double-well shape, leading to the elementary excitation spectrum with a single roton, as shown in Fig.~\ref{Fig5}(a4). In this case, $\chi(q_x)$ and $S(q_x)$  in Fig.~\ref{Fig5}(b4) and (c4) are similar to those of the antiferromagnetic gas, as shown in Fig.~\ref{Fig1}(b2) and (c2).

\begin{figure}[htbp]
	\centering
	\includegraphics[width=15cm]{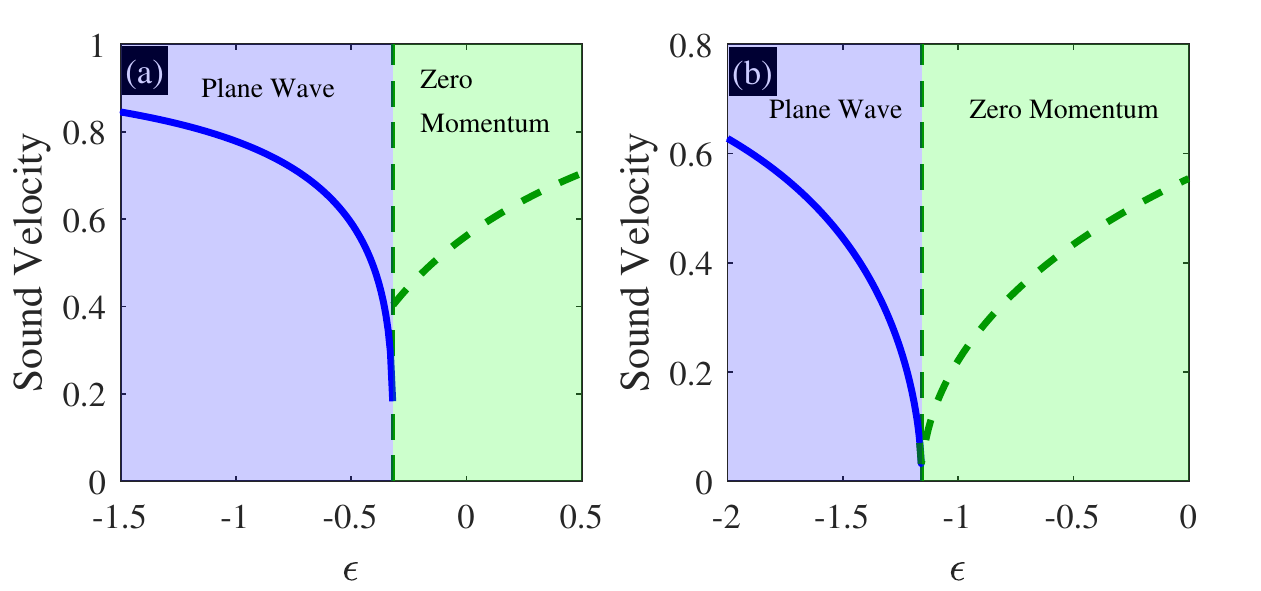}	
	\caption{The sound velocity as a function of the quadratic Zeeman shift $\epsilon$. In (a), $\Omega=1.5$, and In (b), $\Omega=3$. The different colored backgrounds represent different phases of the ground state.}
	\label{Fig6}
\end{figure}

\begin{figure}[htbp]
	\centering
	\includegraphics[width=8cm]{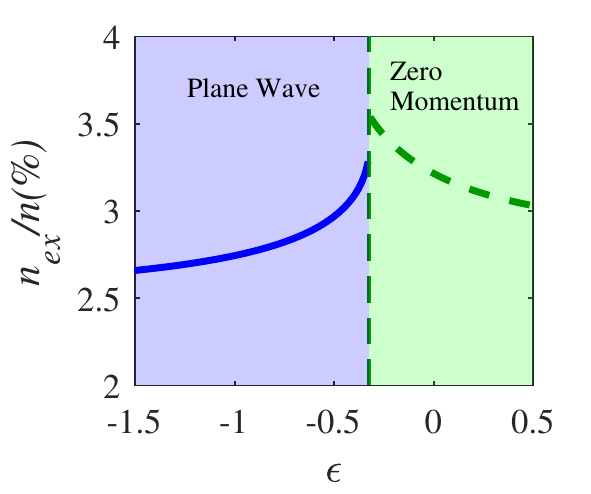}
	\caption{The quantum depletion $n_\mathrm{ex}/n$ as a function of the quadratic Zeeman shift $\epsilon$ with a fixed $\Omega=1.5$.
		The different colored backgrounds represent different phases of the ground state. Here, $g_0n=1$, $g_2n=-0.1$ and $n=1.0k^3_\mathrm{Ram}$.}
	\label{Fig7}
\end{figure}

Once we know the excitation spectrum of the ferromagnetic gas, we can calculate the sound velocity. The sound velocity as a function of the quadratic Zeeman shift $\epsilon$ is demonstrated in Fig.~\ref{Fig6}. Similar to the antiferromagnetic case, the ferromagnetic gas can also undergo first order or second order phase transitions~\cite{Yu}. When $\Omega=1.5$, the ground state experiences a phase transition from the zero-momentum phase to the plane-wave phase around $\epsilon \approx-0.4$. The sound velocity is discontinuous at this critical value, indicating the occurrence of a first order phase transition [see Fig.~\ref{Fig6}(a)]. However, the phase transition is of the second order when we choose a large $\Omega$, as illustrated in Fig.~\ref{Fig6}(b). These results of phase transitions are in accordance with the previous work where the phase transitions are directly determined from the ground-state energy~\cite{Yu}.

We also plot the quantum depletion as a function of the quadratic Zeeman shift in Fig.~\ref{Fig7} [where we choose the 
same parameters as in Fig.~\ref{Fig6}(a)]. Compared with the $g_2>0$ case, the quantum depletion in the case of $g_2<0$ is 
smaller, possibly due to the fact that the negative spin-dependent interaction lowers the system's energy. 
The first order phase transition between the plane-wave phase and the zero-momentum phase is also revealed by the sudden change
in $n_\mathrm{ex}$ across the phase transition, which is in consistence with Fig.~\ref{Fig6}(a).

\section{Conclusion}
\label{conclusion}

In summary, we have systematically studied the elementary excitations in a spin-orbit-coupled spin-1 BEC for both antiferromagnetic and ferromagnetic interactions. We use the static response function and structure factor to characterize the excitation properties of the system. 
We further show that the sound velocity and quantum depletion can be used to identify the phase transitions between three different phases.
Specifically, for ferromagnetic interactions, we reveal rich roton structures in the excitation spectrum of the plane-wave and zero-momentum phases.
For antiferromagnetic interactions, we find that the stripe phase can not only support two gapless modes in the long wave regime, but 
also can support roton excitations at finite momenta. 
The excitation spectrum and structure factors can be measured in experiments by using Bragg spectroscopy
in a $^{87}$Rb BEC for the ferromagnetic case and in a $^{23}$Na BEC for the antiferromagnetic case.
Our results provide new insights in the exploration of the stripe phase in a spinor BEC and thus may be helpful for studying dynamics in such a system.

\section*{Acknowledgment}

This work is supported by the National Natural Science Foundation of China with Grants No. 11974235, No. 11774219 and No. 11974201.
H.L acknowledges the support from the Okinawa Institute of Science and Technology Graduate University.

\end{document}